\documentclass[11pt, a4paper]{article}

\usepackage{amsmath}
\usepackage{amssymb}
\usepackage{amsthm}
\usepackage{graphicx}
\usepackage{natbib}
\usepackage{xcolor}
\usepackage[colorlinks, citecolor=blue, urlcolor=blue]{hyperref}
\usepackage[left = 2.2cm, right = 2.2cm, bottom = 2.5cm, top = 2.5cm]{geometry}
\usepackage{authblk}
\usepackage{tabularray}
\usepackage{tblr-extras}
\UseTblrLibrary{booktabs}
\UseTblrLibrary{siunitx}
\UseTblrLibrary{caption}
\usepackage{xr}
\usepackage{bm}
\usepackage{pdfpages}

\newcommand*{\bb}{\boldsymbol}

\newtheoremstyle{example}
{1em} 
{1em} 
{} 
{} 
{\bfseries} 
{:} 
{.5em} 
{\thmname{#1} \thmnumber{#2}\thmnote{ (#3)}}
\theoremstyle{example}
\newtheorem{example}{Example}

\newtheoremstyle{theorem}
{1em} 
{1em} 
{\itshape} 
{} 
{\bfseries} 
{:} 
{.5em} 
{\thmname{#1} \thmnumber{#2}\thmnote{ (#3)}}

\theoremstyle{theorem}
\newtheorem{theorem}{Theorem}

\theoremstyle{definition}

\DeclareMathOperator{\var}{var}
\DeclareMathOperator{\diag}{diag}
\DeclareMathOperator{\expect}{E}
\DeclareMathOperator{\cum}{cum}

\DeclareMathOperator{\trace}{trace}
\def\einfo{\bb{i}}
\def\iinfo{\bb{V}}
\def\oinfo{\bb{j}}
\def\sabias{b_{(\psi)}}
\def\hatsabias{\hat{b}_{(\psi)}}
\def\sbias{B_{(\psi)}}
\def\hatsbias{\hat{B}_{(\psi)}}
\def\hatvabias{\hat{\bb{b}}_{(\bb\theta)}}
\def\vabias{\bb{b}_{(\bb\theta)}}
\def\dotpsi{\dot{\bb\psi}}
\def\ddotpsi{\ddot{\bb\psi}}
\def\muffin{{\bb \omega}}
\def\btheta{{\bb\theta}}
\def\bbeta{{\bb\beta}}

\def\bx{{\bb x}}

\def\bY{{\bb{Y}}}
\def\by{{\bb{y}}}
\def\bSigma{{\bb\Sigma}}
\def\bmu{{\bb\mu}}
\def\bS{\bb{S}}
\def\b0{\bb{0}}
\def\bA{\bb{A}}
\def\bC{\bb{C}}
\def\bD{\bb{D}}
\def\bE{\bb{E}}
\def\bP{\bb{P}}
\def\bQ{\bb{Q}}

\def\bdelta{\bb{\delta}}
\def\mA{\mathcal{A}}
\def\mS{\mathcal{S}}

\def\bM{\bb{M}}

\def\bd{{\bb{d}}}
\def\bnu{{\bb{\nu}}}
\def\bXi{{\bb{\Xi}}}
\def\bzeta{{\bb{\zeta}}}
\def\bL{{\bb{L}}}
\def\bU{{\bb{U}}}

\author[1,2]{Davide Benussi\thanks{davide.benussi@phd.unipd.it}}
\author[2]{Ioannis Kosmidis\thanks{ioannis.kosmidis@warwick.ac.uk}}
\author[1]{Alessandra Salvan\thanks{alessandra.salvan@unipd.it}}
\author[1]{Nicola Sartori\thanks{nicola.sartori@unipd.it}}

\affil[1]{Department of Statistical Sciences, University of Padova, 35121, Padova, Italy}
\affil[2]{Department of Statistics, University of Warwick, CV4 7AL, Coventry, UK}

\title{Focused median bias reduction}

\begin{document}
\maketitle

\begin{abstract}
  Median bias reduction of maximum likelihood estimators can
  substantially improve estimation and inference. Existing generally
  applicable methods are, however, implicit, requiring the solution of
  nonlinear systems of estimating equations for a specified
  parameterization. Their application to parameter transformations
  often involves tedious algebra and bespoke implementations. We
  develop an explicit median bias-corrected estimator for focus
  parameters defined as smooth scalar transformations of the
  parameter. The estimator results from solving an equation derived
  from the Cornish-Fisher expansion of the centred and scaled maximum
  likelihood estimator of the focus parameter, and requires only the
  maximum likelihood or an asymptotically equivalent estimator of the
  model parameter, the gradient and Hessian of the transformation, and
  expectations of products of log-likelihood derivatives. These
  expectations are available for many models in the bias reduction
  literature and can also be estimated by Monte Carlo simulation. The
  resulting estimators are third-order median unbiased and provide
  one-step approximations to estimators from implicit median bias
  reduction when the focus is a component of the parameter. The method
  can improve standard asymptotic inference and enables hull-based
  confidence procedures to produce intervals with near nominal
  finite-sample coverage. We illustrate the framework with examples
  involving Mahalanobis distances, quantiles, focus parameters in
  regression, stratified, circular, and multiple-mediator models, and
  model selection settings using the Focused Information Criterion.
\end{abstract}

\section{Introduction}
\label{sec:Introduction}

Let $\btheta$ denote the parameter vector of a model, and
$\hat\btheta$ its maximum likelihood (ML) estimator, and consider a
scalar parameter of interest $\psi = h(\btheta)$ for a smooth function
$h(\cdot)$. The ML estimator is $\hat\psi = h(\hat\btheta)$ by
equivariance. Following \citet{claeskens+hjort:2003}, we call $\psi$
the focus parameter to emphasize that different choices of $h(\cdot)$
specify different targets for estimation and inference using the same
model, and that an explicit full parameterization having $\psi$ as a
component does not need to be specified.

When the information about the model parameters is small or moderate
(e.g. small to moderate sample sizes and/or large dimension of
$\btheta$), the finite-sample behaviour of $\hat\psi$ can differ
substantially from that expected by first-order ML theory. These
departures can be particularly pronounced when $h(\cdot)$ is strongly
nonlinear or when $\psi$ is close to a boundary of its range, and
$\hat\psi$ can have appreciable mean bias, substantially different
probabilities of under- and overestimation, and the coverage of
standard Wald-type confidence intervals can be markedly different from
the nominal level. A further difficulty arises when the ML estimates
lie on the boundary of the parameter space, as can occur, for example,
under separation in logistic regression \citep{albert+anderson:1984}.

\citet{firth:1993} proposed solving adjusted score equations to reduce
the asymptotic bias of the ML estimator. In logistic regression with
full rank model matrix, this adjustment yields finite estimates, even
under separation \citep[Corollary~1]{kosmidis+firth:2021}. Mean bias
reduction, however, does not generally carry over under nonlinear
transformations; substituting a mean bias-reduced estimate of
$\btheta$ into $h(\cdot)$ need not give a mean bias-reduced estimate
of $\psi$.  In this direction, \citet{dicaterina+kosmidis:2019}
provide expressions for the leading term in the mean bias expansion of
ML estimator of $h(\btheta)$, with application to Wald transforms.
\citet{daehlen+etal:2024} derive the same expressions from a different
starting point, with applications to focused information criteria.

Median centring, on the other hand, aims to balance the probabilities
of under- and over-estimation and is preserved under strictly monotone
transformations. In addition, it relates to inference more
directly; for example, the coverage guarantees of hull-based
confidence intervals (HulC; \citealp{kuchibhotla+etal:2024}) depend on
a bound of the median bias of the estimators computed within the data
partitions.
One approach to reducing the asymptotic median bias of the ML
estimator is to use the zero of the modified signed likelihood root as
a point estimate \citep{pace+salvan:1999}. The \texttt{likelihoodAsy}
R package \citep{likelihoodAsy} implements an approximation to this
root for inference about scalar focus parameters from general
parametric models.  The implementation involves repeated simulation,
constrained optimization over a grid of focus values, and numerical
inversion. As a result, it may require careful tuning and can be
numerically fragile, particularly with estimates near the
boundary. Another approach is the adjusted score equations of
\citet{pagui+etal:2017}. For a general focus parameter, this requires
either re-expressing the model and the score adjustments in terms of
the focus and a set of nuisance parameters, or deriving and solving an
adjusted profile-score equation. Both routes can be algebraically and
computationally demanding.

We develop an explicit median bias-corrected (MBC) estimator for
scalar focus parameters. The correction is derived from the locally
relevant root of a Cornish-Fisher normalizing transformation of the
centred and scaled ML estimator of the focus parameter. The resulting
estimator is third-order median unbiased under general regularity
conditions, and can be computed using an initial estimate in a chosen
parameterization, the gradient and Hessian of $h(\cdot)$, and
expectations of products of log-likelihood derivatives. When these
expectations are unavailable analytically, they can be approximated by
Monte Carlo simulation. The same initial estimates and the
expectations can be reused across different focus parameters. When the
focus is a component of $\btheta$, the correction starting from ML or
the reduced mean-bias estimator of \citet{firth:1993} coincides with
one quasi-Fisher scoring step towards the implicit median bias-reduced
estimator of \citet{pagui+etal:2017}. HulC intervals constructed using
MBC estimators achieve near nominal finite-sample coverage.

We demonstrate the broad applicability of the framework through
diverse problems, including estimation and inference for squared
Mahalanobis distances, individual marginal effects in generalized
linear models, joint indirect effects in multiple-mediator models, and
risk differences in a model selection setting based on the Focused
Information Criterion (FIC). The Supplementary Material document gives
further results and provides additional examples including
Mahalanobis distances between two distributions, the common shape
parameter in stratified gamma samples, overdispersion in beta-binomial
regression, Weibull quantiles, ordinal superiority measures in
adjacent-category models, and circular variance in a bivariate von
Mises sine model. The accompanying \texttt{focuson} R package
\citep{focuson} provides general implementations of the proposed
correction.


\section{Focused median bias reduction}
\label{sec:fmedbr}

\subsection{Setup and notation}
\label{sec:setup}

Consider a parametric model $\mathcal{M}_\btheta$ with parameter
vector
$\btheta = (\theta_1, \ldots, \theta_p)^\top \in \Theta\subset \Re^p$,
and let $l(\btheta; \by)$ be the corresponding log-likelihood function
at an observed realization
$\bb y = (\by_1^\top, \ldots, \by_k^\top)^\top$ of a random vector
$\bb Y$ distributed according to $\mathcal{M}_\btheta$. The ML
estimator is defined as
$\hat{\btheta} = \arg\max_{\btheta\in \Theta} l(\btheta;
\bY)$. Suppose that $\psi = h(\btheta)$ is a scalar focus parameter,
where $h(\cdot)$ is a smooth function $h: \Theta \to H \subset
\Re$. Inference about $\psi$ can be based on the profile
log-likelihood
$l_p(\psi; \by) = \max_{\btheta \in \Theta: \psi =
  h(\btheta)}l(\btheta; \by)$, giving $\hat\psi = h(\hat{\btheta})$ as
the ML estimator of $\psi$. 
In what follows, all probability statements, expectations, and
asymptotic results are under $\mathcal{M}_\btheta$. We assume that the
log-likelihood function has continuous partial derivatives up to the
required order, and that expectations of products of those derivatives
are finite, again up to the required order. The orders are apparent from
the context. Overall, we assume the standard regularity conditions for
parametric inference, as stated, for example, in \citet[Sections~7.1,
7.2]{mccullagh:2018}.

We also assume the usual regularity conditions for the validity of
Edgeworth and Cornish-Fisher expansions; see, for example,
\citet[Chapter~2]{hall:1992}. All theoretical derivations below
concern the case that $\mathcal{M}_\btheta$ is a continuous
distribution. In the discrete case, the Edgeworth and Cornish-Fisher
expansions we employ also include oscillatory terms \citep[see, for
example,][expression~(A.1)]{cai+wang:2009}. In practice, we ignore
those terms in the derivations. Nevertheless, we illustrate the
effectiveness of the MBC focus estimator in discrete cases, including
individual marginal effects in probit regression
(Example~\ref{ex:marginal-effects}), FIC-driven selection of logistic
regression models (Example~\ref{ex:fic}), and beta-binomial regression
and ordinal superiority measures from ordinal regression models
(Section~\ref{sec:supp:beta-binomial} and
Example~\ref{sec:supp:ordinal-superiority}, respectively, in the
Supplementary Material document).

For convenience of notation, we will be suppressing the argument for
various quantities, after defining the suppression, and restore the
argument when necessary. The vector of $p$ zeros is denoted by
$\b0_p$, $[{\bb K}]_r $ denotes the $r$th row of a matrix ${\bb K}$,
written as a column vector, and $K_{rs}$ its $(r, s)$th
element. Similarly, $L_r$ denotes the $r$th element of a vector
$\bb{L}$.

\subsection{Oracle median bias-corrected focus estimator}
\label{sec:adistr}

Let $\sbias \equiv \sbias(\btheta) = \expect(\hat\psi - \psi)$,
$K_2 \equiv K_2(\btheta) = \var(\hat\psi)$, and
$K_3 \equiv K_3(\btheta) = \cum_3(\hat\psi)$ be the bias, variance and
third cumulant of $\hat\psi$ under $\mathcal{M}_{\btheta}$.  Assuming
that $K_2 > 0$, consider the centred and scaled version of $\hat\psi$,
$W \equiv W(\hat\psi; \btheta) = K_2^{-1/2}\{\hat\psi -
E(\hat\psi)\}$, and let $\rho_3 = K_2^{-3/2} K_3$ be the standardized
third cumulant of $\hat\psi$. A normalizing transformation of $W$
based on a Cornish-Fisher expansion \cite[see, for
example,][equation~(10.20)]{pace+salvan:1997} gives the asymptotically
normal pivot
\begin{equation}
  \label{eq:oracle_cf}
  W - \frac{\rho_3}{6}(W^2 - 1) + O_p(n^{-1}) \, ,
\end{equation}
where $n$ characterizes the rate at which the information about $\btheta$
(and $\psi$) accumulates as the sample size grows, which is typically
but not necessarily equal to $k$. Let
$Z = W - \rho_3(W^2 - 1) /6$ be the asymptotically normal
pivot~\eqref{eq:oracle_cf}, ignoring the $O_p(n^{-1})$ term.

\begin{theorem}[Oracle MBC estimator]
  \label{thm:oracle}
  Assume that $K_2 > 0$, $\rho_3$, and $\sbias$ are known, and let
  $\tilde\psi^{(a)}$ be the solution of the estimating equation
  $Z = 0$ with respect to $\psi$, corresponding to the root with
  $W = O_p(n^{-1/2})$. Then, $\tilde\psi^{(a)}$ is third-order median
  unbiased with $P(\tilde\psi^{(a)} \le \psi) = 1/2 + O(n^{-3/2})$.
\end{theorem}
The proof of Theorem~\ref{thm:oracle} is in
Section~\ref{sec:supp:proof-thm1} of the Supplementary Material
document. Expression~(\ref{eq:oracle-re-expression}) in the proof
suggests that
$\tilde\psi^{(a)} = \hat\psi - \sbias + K_3 / (6 K_2) +
O_p(n^{-3/2})$. Hence, an oracle MBC focus estimator is
\begin{equation}
  \label{eq:oracle_expansion}
  \tilde\psi^{(o)} = \hat\psi - \sbias + \frac{1}{6} \frac{K_3}{K_2} \, .
\end{equation}
The same Edgeworth argument as in the proof gives that
$\tilde\psi^{(o)}$ is third-order median unbiased. In addition,
from~\eqref{eq:oracle_expansion},
$\tilde\psi^{(o)} = \hat\psi + O_p(n^{-1})$. An application of
Slutsky's lemma guarantees that $K_2^{-1/2}(\tilde\psi^{(o)} - \psi )$
converges in distribution to a standard normal, exactly as
$K_2^{-1/2}(\hat\psi - \psi )$ does.

Expression~\eqref{eq:oracle_expansion} requires knowledge of $\sbias$,
$K_2$, and $K_3$, which are, of course, rarely available, both because
they typically depend on $\btheta$ and because the expressions of
$\sbias(\cdot)$, $K_2(\cdot)$, and $K_3(\cdot)$ are typically not
available in closed form.


\subsection{Non-oracle estimators}
\label{sec:focused}

If $\sbias(\cdot)$, $K_2(\cdot)$, and $K_3(\cdot)$ are available in
closed-form, then the dependence on $\btheta$ can be resolved by
replacing $\btheta$ by $\hat\btheta$
in~\eqref{eq:oracle_expansion}. Under the same regularity conditions,
Taylor expansions of $\hatsbias \equiv \sbias(\hat\btheta)$,
$\hat{K}_3 \equiv K_3(\hat\btheta)$ and
$\hat{K}_2 \equiv K_2(\hat\btheta)$ around $\btheta$ give that
\begin{equation}
  - \hatsbias
  + \frac{1}{6}\frac{\hat{K}_3}{\hat{K}_2}
  =
  - \sbias + \frac{1}{6}\frac{K_3}{K_2} + O_p(n^{-3/2}) \, .
\end{equation}
Hence, a non-oracle MBC estimator based on
consistent estimators of $\sbias$, $K_2$ and
$K_3$ is
\begin{equation}
  \label{eq:focused-consistent}
  \tilde\psi
  =
  \hat\psi
  - \hatsbias
  + \frac{1}{6}\frac{\hat{K}_3}{\hat{K}_2} \, .
\end{equation}

\begin{example}[Mahalanobis distance]
  \label{ex:mahalanobis-distance}
  Suppose that $\bY_1, \ldots, \bY_n$ are independent random vectors
  with $\bY_i \sim N_p(\bmu, \bSigma)$, where $N_p(\bmu, \bSigma)$
  denotes a $p$-variate normal distribution with mean $\bmu \in \Re^p$
  and positive-definite $p \times p$ covariance matrix $\bSigma$. The
  squared Mahalanobis distance of a $p$-vector $\bmu_0$ from $\bmu$ is
  $\psi = (\bmu - \bmu_0)^\top \bSigma^{-1} (\bmu - \bmu_0)$. The
  ML estimator of $\psi$ is
  $\hat\psi = (\bar\bY - \bmu_0)^\top \bS^{-1} (\bar\bY - \bmu_0)$,
  where $\bar{\bY} = \sum_{i = 1}^n \bY_i / n$ and
  $\bS = \sum_{i = 1}^n (\bY_i - \bar\bY) (\bY_i - \bar\bY)^\top /
  n$. We can write $\hat\psi = X_1 / X_2$, where $X_1$ is independent
  of $X_2$, with $X_1 \sim {\chi'}^2_p(\lambda)$ and
  $X_2 \sim \chi^2_\nu$, where $\lambda = n \psi$, $\nu = n - p$, and
  ${\chi'}^2_p(\lambda)$ is the noncentral chi-squared distribution
  with non-centrality parameter $\lambda$ and $p$ degrees of freedom,
  and $\chi^2_\nu$ is the chi-squared distribution with $\nu$ degrees
  of freedom. For $\nu_k = n - p - k$ and $\nu_6 > 0$, some
  straightforward algebra, then, gives that
  \begin{align}
    \label{eq:mahalanobis-moments}
    \sbias = \frac{m_1}{\nu_2} - \psi \,, \quad 
    K_2 = \frac{m_2}{\nu_2\nu_4} - \frac{m_1^2}{\nu_2^2} \,, \quad
    K_3 = \frac{m_3}{\nu_2\nu_4\nu_6} - 3 \frac{m_1m_2}{\nu_2^2\nu_4} + 2\frac{m_1^3}{\nu_2^3} \,,
  \end{align}
  where $m_1 = p + \lambda$, $m_2 = m_1 (m_1 + 2) + 2 \lambda$, and
  $m_3 = (m_1^2 + 2m_1)(m_1 + 4) + 6\lambda m_1 + 16 \lambda$. Note
  here that all quantities in~(\ref{eq:mahalanobis-moments}) depend on
  $\bmu$ and $\bSigma$ only through $\psi$. Evaluating those at
  $\hat\psi$, expression~(\ref{eq:focused-consistent}) results in a
  MBC estimator of the squared Mahalanobis distance.

    \begin{table}[t!]
\centering
\begin{talltblr}[         
caption={Comparison of the ML estimator ($\hat\psi$) and the MBC focus estimator ($\tilde\psi$) of the squared Mahalanobis distance in terms of simulation-based estimates (see Example~\ref{ex:mahalanobis-distance}) of mean bias (BIAS), mean absolute deviation (MAD), probability of underestimation (PU), root mean squared error (RMSE), and coverage of $95\%$ Wald-type confidence intervals, where the standard error is estimated by $\{K_2(\hat\psi)\}^{1/2}$ and $\{K_2(\tilde\psi)\}^{1/2}$, respectively. All summaries are $\times 100$.},
label={tab:mahalanobis-distance},
]                     
{                     
colspec={Q[]Q[]Q[]Q[]Q[]Q[]Q[]Q[]},
hline{2}={1-8}{solid, black, 0.05em},
hline{6}={1-8}{solid, black, 0.05em},
hline{10}={1-8}{solid, black, 0.05em},
hline{14}={1-8}{solid, black, 0.05em},
hline{18}={1-8}{solid, black, 0.05em},
hline{22}={1-8}{solid, black, 0.05em},
hline{1}={1-8}{solid, black, 0.08em},
hline{26}={1-8}{solid, black, 0.08em},
column{3}={si={table-format=-4.0,table-align-text-after=false,input-symbols={-,\*+()}},},
cell{1}{3}={guard,halign=c,},
column{4}={si={table-format=-3.1,table-align-text-after=false,input-symbols={-,\*+()}},},
cell{1}{4}={guard,halign=c,},
column{5}={si={table-format=-3.1,table-align-text-after=false,input-symbols={-,\*+()}},},
cell{1}{5}={guard,halign=c,},
column{6}={si={table-format=-2.1,table-align-text-after=false,input-symbols={-,\*+()}},},
cell{1}{6}={guard,halign=c,},
column{7}={si={table-format=-3.1,table-align-text-after=false,input-symbols={-,\*+()}},},
cell{1}{7}={guard,halign=c,},
column{8}={si={table-format=-2.1,table-align-text-after=false,input-symbols={-,\*+()}},},
cell{1}{8}={guard,halign=c,},
row{1}={}{halign=c},
cell{2,10,18}{1}={r=8}{halign=r, valign=h},
cell{2,6,10,14,18,22}{2}={r=4}{halign=l, valign=h},
cell{3-5,7-9,11-13,15-17,19-21,23-25}{2}={}{halign=l},
cell{3-9,11-17,19-25}{1}={}{halign=r},
}                     
$p$ & Estimator & $n$ & BIAS & MAD & PU & RMSE & Coverage \\
10 & $\hat\psi$ & 128 & 26.0 & 34.5 & 24.7 & 44.7 & 96.6 \\
10 & $\hat\psi$ & 256 & 12.3 & 20.8 & 31.6 & 26.7 & 95.6 \\
10 & $\hat\psi$ & 512 & 6.0 & 13.5 & 36.8 & 17.2 & 95.3 \\
10 & $\hat\psi$ & 1024 & 3.0 & 9.2 & 40.5 & 11.6 & 95.1 \\
10 & $\tilde\psi$ & 128 & 1.0 & 26.1 & 52.3 & 33.2 & 96.6 \\
10 & $\tilde\psi$ & 256 & 0.9 & 18.0 & 50.7 & 22.7 & 96.0 \\
10 & $\tilde\psi$ & 512 & 0.6 & 12.6 & 50.3 & 15.8 & 95.6 \\
10 & $\tilde\psi$ & 1024 & 0.3 & 8.8 & 50.1 & 11.1 & 95.3 \\
20 & $\hat\psi$ & 128 & 76.1 & 79.7 & 8.9 & 97.4 & 94.3 \\
20 & $\hat\psi$ & 256 & 34.5 & 40.2 & 17.4 & 50.5 & 92.7 \\
20 & $\hat\psi$ & 512 & 16.4 & 23.1 & 25.6 & 29.3 & 93.4 \\
20 & $\hat\psi$ & 1024 & 8.0 & 14.6 & 32.1 & 18.4 & 94.1 \\
20 & $\tilde\psi$ & 128 & -8.8 & 40.1 & 60.7 & 49.9 & 96.7 \\
20 & $\tilde\psi$ & 256 & -0.7 & 26.7 & 53.3 & 33.6 & 96.5 \\
20 & $\tilde\psi$ & 512 & 0.3 & 18.5 & 51.1 & 23.2 & 95.9 \\
20 & $\tilde\psi$ & 1024 & 0.3 & 12.9 & 50.4 & 16.2 & 95.5 \\
30 & $\hat\psi$ & 128 & 160.0 & 160.9 & 2.2 & 185.0 & 91.8 \\
30 & $\hat\psi$ & 256 & 68.5 & 71.5 & 8.1 & 85.7 & 87.7 \\
30 & $\hat\psi$ & 512 & 32.0 & 37.0 & 16.4 & 45.9 & 89.8 \\
30 & $\hat\psi$ & 1024 & 15.5 & 21.5 & 24.6 & 26.9 & 92.0 \\
30 & $\tilde\psi$ & 128 & -40.8 & 62.9 & 76.1 & 75.6 & 94.2 \\
30 & $\tilde\psi$ & 256 & -6.0 & 36.0 & 57.8 & 45.0 & 96.7 \\
30 & $\tilde\psi$ & 512 & -0.6 & 24.5 & 52.5 & 30.8 & 96.2 \\
30 & $\tilde\psi$ & 1024 & 0.2 & 17.0 & 50.8 & 21.4 & 95.7 \\
\end{talltblr}
\end{table}

    Table~\ref{tab:mahalanobis-distance} compares the performance of
    $\hat\psi$ and $\tilde\psi$, and of corresponding Wald-type
    confidence intervals, by simulating $10^6$ samples for each
    combination of $n = 2^q$, $q \in \{7, 8, 9, 10\}$, and
    $p \in \{10, 20, 30\}$ from a $N_p(\bmu, \bSigma)$ model with
    $\bmu = (1/p, 2/p, \ldots, 1)^\top$ and
    $\bSigma_{ij} = 1/2^{|i - j|}$. The MBC estimator outperforms the
    ML one with probability of underestimation substantially closer to
    $50\%$ in all cases, also delivering substantial improvements in
    terms of finite-sample mean bias, mean absolute deviation and root
    mean squared error. As $p$ increases, Wald-type intervals based on
    $\tilde\psi$ tend to have coverage substantially closer to the
    $95\%$ nominal level than those based on $\hat\psi$.
\end{example}

Similar improvements are noted in the estimation and inference about
the squared Mahalanobis distance between two distributions; see
Section~\ref{sec:supp:mahalanobis-distance-two-sample} of
the Supplementary Material document.

In cases where $\sbias$, $K_2$, and $K_3$ are not available in closed
form, we can approximate them using suitable asymptotic expansions,
and, again, replace $\btheta$ with $\hat{\btheta}$. From the results
in \citet[Section 4.3]{kosmidis+firth:2010}, the bias of $\hat\psi$ is
written as
\begin{equation}
  \label{eq:mle_phi_bias}
  \sbias = b_{(\psi)} + O(n^{-2})\,, \quad \sabias = \vabias^\top\dotpsi +
  \frac{1}{2} \trace(\iinfo \ddotpsi) \, ,
\end{equation}
where $\vabias \equiv \vabias(\btheta)$ is the leading term in the
bias expansion of $\hat\btheta$, and
$\dotpsi \equiv \dotpsi(\btheta) = \nabla h(\btheta)$ and
$\ddotpsi \equiv \ddotpsi(\btheta) = \nabla \nabla^\top h(\btheta)$
are the gradient and the Hessian of $h(\btheta)$,
respectively. Furthermore,
$\einfo \equiv \einfo(\btheta) = \expect(\oinfo)$ is the expected
information matrix, where
$\oinfo \equiv \oinfo(\btheta; \bY) = -\nabla \nabla^\top l(\btheta;
\bY)$ is the observed information matrix at $\btheta$, and
$\iinfo = \einfo^{-1}$. The first term in the bias expansion of
$\hat\btheta$ has the form $\vabias = -\iinfo \bA$ \citep[see, for
example,][Section~2]{kosmidis+firth:2010} where the $t$th element of
$\bA$ is $A_t = \trace\{\iinfo (\bP_t + \bQ_t)\} / 2$
$(t = 1, \ldots, p)$. In the latter expression,
$\bP_t \equiv \bP_t(\btheta) = \expect(\bb{s} \bb{s}^\top s_t)$ and
$\bQ_t \equiv \bQ_t(\btheta) = -\expect(\bb{j} s_t)$, where
$\bb{s} \equiv \bb{s}(\btheta; \bb Y) = \nabla l(\btheta; \bb Y)$ and
$s_t$ is the $t$th element of $\bb{s}$. For the variance of
$\hat\psi$, a delta-method argument gives that
\begin{equation}
  \label{eq:mle_phi_variance}
  K_2 = \kappa_2 + O(n^{-2}) \,, \quad \kappa_2 = \muffin^\top \dotpsi \, ,
\end{equation}
where $\muffin = \iinfo \dotpsi$. Finally, in
Section~\ref{sec:supp:k3} of the Supplementary Material document we show that
\begin{equation}
  \label{eq:cum_phi}
K_3 = \kappa_3 + O(n^{-3}) \,, \quad \kappa_3 = -6 \sum_{t = 1}^p \omega_t \muffin^\top \left(\frac{1}{3} \bP_t + \frac{1}{2} \bQ_t\right) \muffin + 3 \muffin^\top \ddotpsi \muffin \, ,
\end{equation}
where $\omega_t$ is the $t$th element of
$\muffin$. From~\eqref{eq:mle_phi_bias},~\eqref{eq:mle_phi_variance},~\eqref{eq:cum_phi},
assuming that $n\kappa_2$ is bounded away from zero, Taylor expansions
of $\hatsabias \equiv \sabias(\hat\btheta)$,
$\hat{\kappa}_3 \equiv \kappa_3(\hat\btheta)$ and
$\hat{\kappa}_2 \equiv \kappa_2(\hat\btheta)$ around $\btheta$ can be
used to show that the plug-in correction satisfies
\[
  - \hatsabias
  + \frac{1}{6}\frac{\hat{\kappa}_3}{\hat{\kappa}_2}
  =
  -\sbias
  + \frac{1}{6}\frac{K_3}{K_2}
  + O_p(n^{-3/2}) \, .
\]
Hence, a non-oracle MBC estimator based on
first-order approximations of $\sbias$, $K_2$ and $K_3$ is
\begin{equation}
  \label{eq:focused}
  \tilde\psi
  =
  \hat\psi
  - \hatsabias
  + \frac{1}{6}\frac{\hat{\kappa}_3}{\hat{\kappa}_2} \, .
\end{equation}

The estimators in~\eqref{eq:focused-consistent} and~\eqref{eq:focused}
agree with~\eqref{eq:oracle_expansion} up to $O_p(n^{-3/2})$.  Under
the stated regularity conditions for continuous models, higher-order
delta-method arguments \citep[see][Section~2.7]{hall:1992} establish
that those estimators are also third-order median unbiased.

The expected information matrix, and the expected values of products
of log-likelihood derivatives $\bP_t$ and $\bQ_t$ are available for a
wide range of statistical models where mean and median bias reduction
techniques have been previously implemented. A few instances include
\citet{kosmidis+etal:2020} for generalized linear models;
\citet{kosmidis+firth:2011} and \citet{kosmidis:2014a} for baseline
category multinomial regression and cumulative link models for ordinal
responses, respectively; \citet{grun+etal:2012} for beta regression;
\citet{pozza+etal:2023} for relative risk regression;
\citet{pagui+etal:2022} for negative binomial regression models. More
generally, they have also implicitly or explicitly been derived for
models where the first-order bias term $\vabias(\btheta)$ has been
computed through the general expression provided in
\citet{cox+snell:1968}.  The MBC estimator for focus
parameters is readily available or straightforward to derive in all
those cases.

If $\mathcal{M}_\btheta$ is a full exponential family in canonical
parameterization, then the algebraic effort for
deriving~\eqref{eq:focused} is reduced substantially. In that case
$\oinfo(\btheta; \bY) = \oinfo(\btheta) = \einfo(\btheta)$, and,
hence, $\bQ_t$ is a $p \times p$ matrix of zeros. Also,
$\bP_1, \ldots, \bP_p$ are matrices of third-order cumulants of the
sufficient statistics. Such an example is
\citet[Example~5.3]{pace+salvan:1999}, where the focus parameter is
the common shape parameter in stratified gamma samples. For that
particular setting, Section~\ref{sec:supp:gamma-shape} of the
Supplementary Material document shows that the MBC focus estimator
coincides with the approximate optimal median-unbiased estimator that
has been derived in \citet[Example~5.3]{pace+salvan:1999} by finding
the zero of the modified signed likelihood root (see
Section~\ref{sec:rstar}).

In expressions~(\ref{eq:focused-consistent}) and~(\ref{eq:focused}),
instead of $\hat\btheta$, we can use any other consistent estimator
$\btheta^\dagger$ of $\btheta$ satisfying
$\btheta^\dagger = \hat\btheta + n^{-1}\bd(\btheta) + O_p(n^{-3/2})$,
for some sufficiently smooth vector-valued function $\bd(\btheta)$,
with the
caveat that $\sbias$ and $\sabias$ are now the bias and approximate
bias, respectively, of $\psi^\dagger= h(\btheta^\dagger)$. Especially
for~(\ref{eq:focused}), using an estimator with $o(n^{-1})$ mean bias
is useful because then $\sabias = \trace(\iinfo \ddotpsi) / 2$. Such
an estimator can be selected strategically in light of the
side-effects that mean bias reduction may have for certain models. For
example, \citet[Corollary~1]{kosmidis+firth:2021} shows that the
reduced-bias estimator of \citet{firth:1993}, which results implicitly
as the solution of $\bb{s} + \bA = \b0_p$ with respect to $\btheta$,
always takes finite values in logistic regression with full rank model
matrices. Also, \citet[Section~5.3]{kosmidis+lunardon:2024} discuss
how reduced-bias estimators that are always in the interior of the
parameter space can be constructed for more general models using extra
plug-in penalties to a bias-reducing penalized log-likelihood. Another
option is the maximum-softly penalized likelihood estimator of
\citet{sterzinger+kosmidis:2023} that has been developed for logistic
mixed effects models but applies more generally. That estimator has
been shown to avoid boundary estimates, while preserving the same
asymptotic bias and asymptotic distribution as the ML
estimator.

\begin{example}[Individual marginal effects for generalized linear models]
  \label{ex:marginal-effects}
  Suppose that $y_1, \ldots, y_k$ are realisations of independent
  random variables $Y_1, \ldots, Y_k$, each with probability density
  or mass function of the exponential family form
  \begin{equation*}
    f(y; \zeta_i, \phi) = \exp \left\{ \frac{y\zeta_i - b(\zeta_i) - c_1(y)}{\phi / m_i} -
    \frac{1}{2} a\left(-\frac{m_i}{\phi}\right) + c_2(y)\right\} \,,
\end{equation*}
for sufficiently smooth functions $b(\cdot)$, $c_1(\cdot)$,
$a(\cdot)$, and $c_2(\cdot)$, and known weights $m_1, \ldots,
m_k$. Hence, $\expect(Y_i) = \mu_i = b'(\zeta_i)$ and
$\var(Y_i) = \phi V(\mu_i) / m_i$, with $V(\mu_i) = b''(\zeta_i)$. A
generalized linear model links $\mu_i$ to a linear predictor $\eta_i$
through a monotone, sufficiently smooth link function
$g(\mu_i) = \eta_i$ with $\eta_i = \sum_{t = 1}^p \beta_t x_{it}$,
where $x_{it}$ is the $(i, t)$th component of a model matrix $X$, and
$\bbeta = (\beta_1, \ldots, \beta_p)^\top$. The individual marginal
effect of a continuous covariate $z$ at covariate setting $\bx$ is
$\psi = d \dot{\eta}$, where
$d \equiv d(\eta)= d g^{-1}(\eta) / d\eta$, $\eta = \bx^\top \bbeta$,
and $\dot{\eta} = \dot{\bx}^\top \bbeta$ with
$\dot{\bx} = \partial \bx / \partial z$. The definition of the
marginal effect accounts for the case where more than one element of
the covariate vector $\bx$ may be functions of the covariate $z$, such
as when interaction terms, polynomials of a single covariate or other
nonlinear functions of covariates are included in the covariate vector
$\bx$. Estimation and inference about marginal effects is typically
carried out using $\hat\bbeta$ in place of $\bbeta$ in the definition
of $\psi$, and the delta method; see, for example,
\citet{arel-bundock+etal:2024} for the \texttt{marginaleffects} R
package.

\begin{figure}[t]
  \centering
  \includegraphics[width = \linewidth]{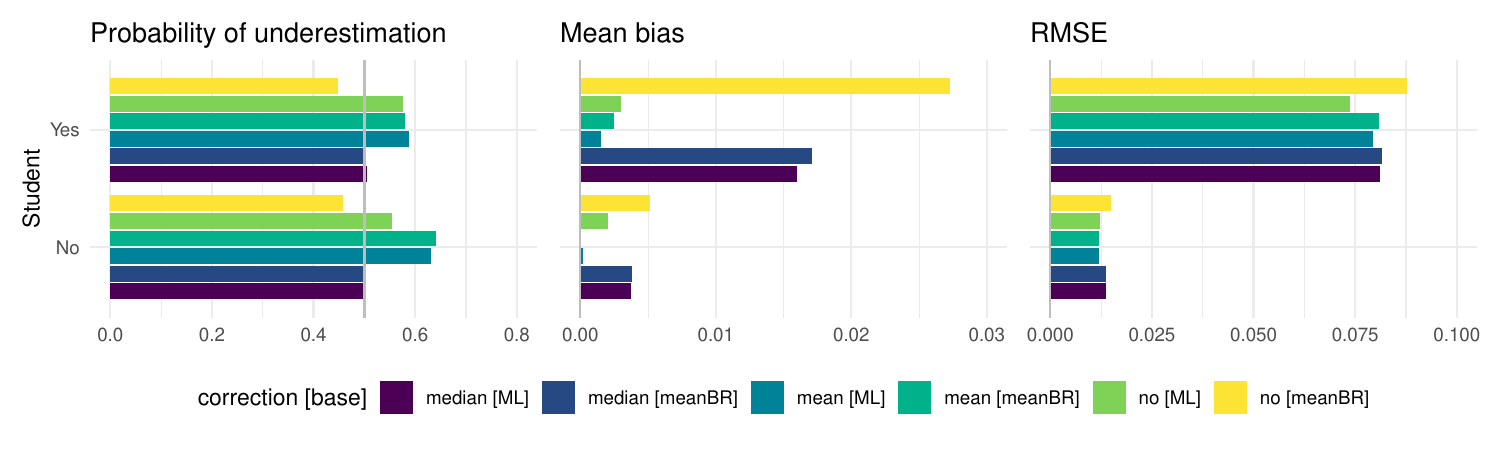}
  \caption{Comparison of various estimators of two individual marginal
    effects in a probit regression model, evaluated at covariate
    settings with $\texttt{Student}=\texttt{No}$ and
    $\texttt{Student}=\texttt{Yes}$.  The panels report
    simulation-based estimates (see Example~\ref{ex:marginal-effects})
    of probability of underestimation, mean bias, and root mean squared
    error (RMSE).  Results are shown for the uncorrected plug-in
    estimator (``no''), the mean bias-corrected focus estimator
    (``mean''), and the median bias-corrected focus estimator
    (``median''). The label in brackets indicates whether the plug-in
    quantities are evaluated at the maximum likelihood (ML) or the
    reduced mean-bias (meanBR) estimators of $\bbeta$
    \citep{firth:1993}.}
  \label{fig:marginal-effects}
\end{figure}

Computing the MBC estimator~(\ref{eq:focused}) of
individual marginal effects for any generalized linear model is
straightforward. The matrices $\einfo(\btheta)$, $\bP_t(\btheta)$ and
$\bQ_t(\btheta)$ for generalized linear models, where
$\btheta = (\bbeta^\top, \phi)^\top$, have been derived by
\citet{kosmidis+etal:2020}, and are implemented in the
\texttt{enrichwith} R package \citep{enrichwith}. The gradient and the
Hessian of
$\psi$ with respect to
$\bbeta$ are
\begin{equation}
  \label{eq:me-ders}
  \dotpsi_{(\bbeta)} = d \dot{\bx} + d' \bx \dot{\eta}\,,
  \quad 
  \ddotpsi_{(\bbeta)} = d'(\dot{\bx} \bx^\top + \bx \dot{\bx}^\top) + \dot{\eta} d'' \bx \bx^\top\,,
\end{equation}
respectively, where $d' \equiv d'(\eta) = d^2 g^{-1}(\eta) / d\eta^2$
and $d'' \equiv d''(\eta) = d^3 g^{-1}(\eta) / d\eta^3$. Furthermore,
for models with unknown dispersion, the first and second derivatives
of the marginal effect with respect to $\phi$ are both zero, as are
the second derivatives of $\psi$ with respect to $\beta_t$ and $\phi$
$(t = 1, \ldots, p)$. Using the above, we can compute $\tilde\psi$ by
computing $\hat{\bbeta}$ (and the ML estimator
$\hat{\phi}$ of $\phi$, if required), $\hat\psi$, and $\hatsabias$,
$\hat{\kappa}_2$, and $\hat{\kappa}_3$ and plugging those
in~(\ref{eq:focused}).

As a demonstration, we consider the Default data set that is
provided in the \texttt{ISLR2} R package \citep{ISLR2}. That simulated
data set contains information about whether a customer has defaulted
or not (Yes or No), whether the customer is a student or not (Yes or
No), the average balance the customer has remaining on their credit
card after making their monthly payment (in tens of thousands of
dollars), and the income of the customer (in tens of thousands of
dollars). For a randomly selected subset of $2500$ customers, we model
the probability of default using a probit regression model with an
intercept, main effects for student status, balance and income, and
the interactions of student status with balance and income. Interest
is in the individual marginal effects of balance on the probability of
default at the covariate settings with average income and average
balance for students ($1.797$ and $0.100$, respectively) and
non-students ($3.991$ and $0.077$, respectively). We simulate $10^4$
samples from the ML fit, and compute the uncorrected, and the mean and
MBC estimators of each of the two individual marginal effects, based
on the ML estimator of $\bbeta$ and on the reduced-bias estimators of
\citet{firth:1993} of $\bbeta$ as computed by the \texttt{brglm2} R
package \citep{brglm2}. The mean bias-corrected estimators are
computed as $\hat\psi - \hatsabias$ when based on $\hat\bbeta$, and as
$\psi^\dagger - \sabias^\dagger$ when based on the reduced-bias
estimator $\bbeta^\dagger$. No infinite estimates were
encountered. Figure~\ref{fig:marginal-effects} shows the results. The
MBC estimators have almost completely corrected the median bias of the
uncorrected estimators. The substantial correction in median bias,
here, is accompanied by an inflation in mean bias, which results in a
slight inflation of the root mean squared error.

\end{example}


\section{Explicit and implicit median bias reduction}
\label{sec:exim}

\subsection{Equivariance}
\label{sec:equivariance}

The non-oracle MBC estimators
stemming from Theorem~\ref{thm:oracle} are explicit in the sense that
estimation proceeds by adding a data-dependent adjustment to an
initial estimate. An alternative median bias reduction method has been
proposed in \citet{pagui+etal:2017} that operates implicitly by
solving a system of adjusted score equations
\begin{equation}
  \label{eq:imedian}
  \bb{s} + \bA - \einfo \bC = \b0_p
\end{equation}
with respect to $\btheta$, where $\bC$ has $r$th element
$C_r = [\iinfo]_r^\top [\bD]_r$ with the $p \times p$ matrix $\bD$
having $(r, s)$th component
$D_{rs} = \trace\{ \bE_r (\bP_s / 3 + \bQ_s / 2)\}$ with
$\bE_r = [\iinfo]_r[\iinfo]_r^\top /
V_{rr}$. Equation~(\ref{eq:imedian}) results by stitching together the
$p$ median bias-reducing adjusted profile score equations that result
by considering each element of $\btheta$ as being of interest and all
others as nuisance, and avoiding the evaluation of constrained
estimates; see \citet[Section~2.2 and Section~3]{pagui+etal:2017} for
details. When it exists, the solution of~(\ref{eq:imedian}) is
componentwise median unbiased to third order, and exactly equivariant
under monotone transformations acting separately on each element of
$\btheta$. The MBC estimators stemming from Theorem~\ref{thm:oracle},
on the other hand, are not exactly equivariant under monotone
transformations of the focus parameter, in the sense that the MBC
estimators of $g(\psi)$ for a monotone scalar function $g(\cdot)$ are
not exactly equal to $g(\tilde\psi)$. This is because we use a
specific parameterization for constructing $\tilde\psi$.
Nevertheless, since the events $\{\tilde\psi < \psi\}$ (or
$\{\tilde\psi > \psi\}$) and $\{g(\tilde\psi) < g(\psi)\}$ have equal
probability for increasing (or decreasing) $g(\cdot)$, $g(\tilde\psi)$
is a third-order median unbiased estimator of
$g(\psi)$. Section~\ref{sec:supp:beta-binomial} of the Supplementary
Material document, illustrates the behaviour of the MBC estimator
under reparameterization with a beta-binomial regression example.

\subsection{Median bias-correction at the chosen parameterization}
\label{sec:ex+im}
Importantly, and as is also clear from~(\ref{eq:imedian}), the
implicit median bias reduction method of \citet{pagui+etal:2017}
requires a fully specified nuisance parameterization. Applying it to
arbitrary focus parameters can be both algebraically demanding and
computationally complicated in general models. The ingredients
in~(\ref{eq:imedian}) need to be re-expressed in terms of the focus
parameter, or, alternatively, the median bias-reducing adjusted
profile score equation will need to be derived
\citep[see][expression~(8)]{pagui+etal:2017} and solved, which would
require repeated constrained optimization for the ML
estimators of the nuisance parameters for given values of the focus
parameter.

Despite their very different starting points, there exists a strong
connection between the explicit MBC estimator of
$\psi$ and implicit median bias reduction in the special case where
the focus parameter $\psi$ is one component of $\btheta$. Suppose that
$\psi = \theta_t$. Then, $\dotpsi$ is the $t$th standard basis vector
and $\ddotpsi = \b0_{p \times p}$,
$\hatsabias = [\hatvabias]_t = -[\hat\iinfo]_t^\top \hat{\bA}$, and
$\hat{\kappa_2} = \hat{V}_{tt}$, $\hat{\muffin} =
[\hat{\iinfo}]_t$. Hence, from~(\ref{eq:cum_phi}), and plugging
into~(\ref{eq:focused}), the MBC estimator of
$\theta_t$ is
$\hat\theta_t + [\hat\iinfo]_t^\top \hat{\bA} - \hat{C}_t$
$(t = 1, \ldots, p)$, or in vector form
$\hat\btheta + \hat{\iinfo} \hat{\bA} - \hat{\bC}$, which is exactly
one step of the quasi-Fisher scoring iteration
$\btheta := \btheta + \iinfo (\bb{s} + \bA - \einfo \bC)$ for
solving~(\ref{eq:imedian}) starting from $\hat\btheta$, since
$\hat{\bb{s}} = \b0_p$. Similarly, if the base estimator is the
reduced mean-bias estimator $\tilde\btheta$ of \citet{firth:1993} (see
Example~\ref{ex:marginal-effects}), then the MBC
estimator of the elements of $\btheta$ is
$\tilde\btheta - \tilde{\bC}$, which is one step of the same
quasi-Fisher scoring iteration starting from $\tilde\btheta$, since
$\tilde{\bb{s}} + \tilde{\bA} = \b0_p$. So, the MBC
estimator is a one-step approximation to the estimator from implicit
median bias reduction.

As a result, although the explicit and implicit approaches differ in
their equivariance properties and in the estimators they produce, they
deliver the same third-order median unbiasedness. The explicit MBC
estimator is therefore a natural candidate for a default method when
interest is in scalar focus parameters: it accommodates arbitrary
focus parameters, requiring only their gradient and Hessian with
respect to the parameter; it allows the choice of both
parameterization and initial estimator; and it avoids further
optimization or repeated constrained optimization, beyond the
computation of the initial estimates.


\section{Inference}

\label{sec:infer}

\subsection{Wald-type inference}
\label{sec:wald-inference}

The asymptotic normality result in Section~\ref{sec:adistr} enables
Wald-type inferential procedures to apply by simply replacing
$\hat\psi$ with $\tilde\psi$ and using a consistent estimator of $K_2$
or $\kappa_2$. In Example~\ref{ex:mahalanobis-distance}, we
illustrated that such inferential procedures based on $\tilde\psi$
perform substantially better than those based on $\hat\psi$ for
inference about the squared Mahalanobis distance. However, their
validity is only asymptotic. In fact, there are no general guarantees
that such procedures will perform asymptotically better than those
based on $\hat\psi$. For example, Edgeworth expansions can be used to
show that the two-sided symmetric $100(1- \alpha)\%$ confidence
intervals from the inversion of the asymptotically normal pivots
$K_2^{-1/2} (\hat\psi - \psi)$ and $K_2^{-1/2} (\tilde\psi - \psi)$,
with oracle $K_2$, both have coverage $1 - \alpha + O(n^{-1})$.

For general models, $K_2$ or $\kappa_2$ typically depend on the
parameter vector $\btheta$, while median bias-correction only delivers
an estimate for $\psi$. In practice, for Wald-type procedures such as
confidence intervals, we can simply evaluate $K_2$ or $\kappa_2$ at
the initial estimates of $\btheta$. An alternative is to construct a
parameter vector that is compatible with the corrected focus estimate,
similarly to how Wald-type inferences operate with ML, and implicit
mean and median bias reduction methods (see, for example,
\citealt[Section~5]{firth:1993} and
\citealt[Section~2.2]{pagui+etal:2017}). Section~\ref{sec:supp:compatible-ses}
of the Supplementary Material document describes a procedure to
construct a compatible parameter vector, starting from componentwise
median bias-corrected estimates of $\btheta$ and adjusting one
component so that the transformed estimate agrees as closely as
possible with $\tilde\psi$.

\subsection{Hull-based confidence intervals}
\label{sec:hulc}

The HulC method \citep{kuchibhotla+etal:2024} constructs
confidence intervals by splitting the sample into independent batches,
computing an estimator on each batch, and taking the interval between
the smallest and largest resulting estimates. The coverage guarantees
that HulC provides depend on a uniform upper bound $\Delta$ on the
median bias of the batch estimators and do not require variance
estimation or an explicit limiting distribution. When the batch
estimators are third-order median unbiased, as when~\eqref{eq:focused}
is applied within each batch, then the resulting HulC interval, after
calibration with $\Delta = 0$ and randomization on the number of
batches, achieves sixth-order accuracy
\citep[Section~2.2]{kuchibhotla+etal:2024}. This is remarkable in
terms of inferential guarantees, stemming merely from repeated
calculation of MBC estimators of focus parameters on a few partitions
of data; see Section~\ref{sec:supp:hulc} of the Supplementary Material
document for details.

\begin{example}[Joint indirect effect with multiple mediators]
  \label{ex:mediation-analysis}
  
  For subject $i \in \{1,\ldots,n\}$, let $A_i\in\{0,1\}$ denote a
  randomized treatment, $\bx_i\in \Re^p$ the available baseline
  covariates, $\bM_i\in\Re^q$ the candidate mediators, and $Y_i$ a
  continuous outcome. For the purposes of this example, we consider a
  multiple-mediator model \citep[see, for
  example][Section~2]{vanderweele+vansteelandt:2013}
  \begin{equation}
    \label{eq:mediation-model}
    Y_i = \alpha + \bbeta^\top \bM_i + \gamma A_i + \bdelta^\top \bx_i + \epsilon_{i} \,, \quad
    \bM_i = \bmu + \bnu A_i + \bXi \bx_i + \bzeta_i \,,
  \end{equation}
  where $\epsilon_{i}\sim N(0,\sigma^2)$ and
  $\bzeta_{i}= \bL \bD \bU_i$, where
  $\bD = \diag\{e^{\eta_1}, \ldots, e^{\eta_q}\}$ and the components
  of $\bU_i$ are independent, standardized skew-normal random
  variables with common shape $\lambda \ge 0$.  We also assume that
  $\bU_i$ and $\epsilon_{i}$ are independent, independent across
  subjects, and independent of the treatment assignments and baseline
  covariates.  Thus,
  $\bSigma_M = \var(\bzeta_i) = \bL \bD^2 \bL^\top$. Under
  consistency, positivity, and usual sequential ignorability
  assumptions for multiple-mediator models \citep[see, for
  instance][Section 3.4]{jones+al:2025}, $\psi = \bnu^\top \bbeta$ is
  the joint indirect effect, which is the treatment effect transmitted
  jointly through the mediators \citep[see, for
  example,][]{robins+greenland:1992}.
  Section~\ref{sec:supp-multiple-mediator} in the Supplementary
  Material document derives the leading terms of the mean bias,
  variance and third cumulant of the ML estimator $\hat\psi$ of
  $\psi$, for fixed $\lambda$ and $\bL$. Evaluating those at the ML
  estimator, expression~\eqref{eq:focused} yields the MBC estimator of
  $\psi$.

\begin{table}[t!]
\centering
\begin{talltblr}[         
caption={Comparison of the ML focus estimator ($\hat\psi$) and the MBC focus estimator ($\tilde\psi$) of the joint indirect effect for $\tau = 1$ and $n_1 =n / 16$, in terms of simulation-based estimates (see Example~\ref{ex:mediation-analysis}) of probability of underestimation (PU), mean bias (BIAS), and root mean squared error (RMSE). The column ``Wald'' reports the coverage of nominally $95\%$ Wald-type intervals using the estimated standard error evaluated at the supplied maximum likelihood estimate of $\btheta$. The column ``HulC'' reports the coverage of nominally $95\%$ HulC-type intervals with $\Delta = 0$, constructed using the corresponding estimator. All summaries are $\times 100$.},
label={tab:multiple-mediator-c1},
]                     
{                     
colspec={Q[]Q[]Q[]Q[]Q[]Q[]Q[]Q[]},
hline{2}={1-8}{solid, black, 0.05em},
hline{5}={2-8}{solid, black, 0.05em},
hline{8}={1-8}{solid, black, 0.05em},
hline{11}={2-8}{solid, black, 0.05em},
hline{14}={1-8}{solid, black, 0.05em},
hline{17}={2-8}{solid, black, 0.05em},
hline{1}={1-8}{solid, black, 0.08em},
hline{20}={1-8}{solid, black, 0.08em},
row{1}={}{halign=c},
cell{2-19}{3}={}{halign=r},
cell{2-19}{4}={}{halign=r},
cell{2-19}{5}={}{halign=r},
cell{2-19}{6}={}{halign=r},
cell{2-19}{7}={}{halign=r},
cell{2-19}{8}={}{halign=r},
cell{2,5,8,11,14,17}{2}={r=3}{halign=l, valign=h},
cell{2,8,14}{1}={r=6}{halign=r, valign=h},
cell{3-4,6-7,9-10,12-13,15-16,18-19}{2}={}{halign=l},
cell{3-7,9-13,15-19}{1}={}{halign=r},
}                     
$q$ & Estimator & $n$ & PU & BIAS & RMSE & Wald & HulC \\
10 & $\hat\psi$ & 256 & 39.0 & 0.9 & 2.9 & 97.5 & 90.3 \\
10 & $\hat\psi$ & 512 & 41.2 & 0.4 & 1.7 & 97.2 & 90.3 \\
10 & $\hat\psi$ & 1024 & 43.4 & 0.2 & 1.1 & 96.4 & 91.5 \\
10 & $\tilde\psi$ & 256 & 49.4 & 0.2 & 2.8 & 98.0 & 94.4 \\
10 & $\tilde\psi$ & 512 & 50.0 & 0.1 & 1.6 & 97.3 & 94.8 \\
10 & $\tilde\psi$ & 1024 & 50.2 & 0.0 & 1.1 & 96.4 & 95.0 \\
20 & $\hat\psi$ & 256 & 39.4 & 1.0 & 3.5 & 97.7 & 92.9 \\
20 & $\hat\psi$ & 512 & 40.6 & 0.5 & 1.9 & 97.8 & 92.1 \\
20 & $\hat\psi$ & 1024 & 42.7 & 0.2 & 1.1 & 97.3 & 92.3 \\
20 & $\tilde\psi$ & 256 & 49.2 & 0.2 & 3.2 & 98.5 & 94.7 \\
20 & $\tilde\psi$ & 512 & 49.7 & 0.1 & 1.8 & 98.3 & 94.9 \\
20 & $\tilde\psi$ & 1024 & 50.0 & 0.0 & 1.1 & 97.5 & 95.0 \\
30 & $\hat\psi$ & 256 & 40.7 & 1.0 & 4.1 & 97.6 & 94.1 \\
30 & $\hat\psi$ & 512 & 41.4 & 0.5 & 2.1 & 97.9 & 93.0 \\
30 & $\hat\psi$ & 1024 & 43.0 & 0.2 & 1.2 & 97.6 & 93.0 \\
30 & $\tilde\psi$ & 256 & 49.4 & 0.1 & 3.8 & 98.4 & 94.8 \\
30 & $\tilde\psi$ & 512 & 49.6 & 0.1 & 2.0 & 98.5 & 94.9 \\
30 & $\tilde\psi$ & 1024 & 50.1 & 0.0 & 1.1 & 98.0 & 95.1 \\
\end{talltblr}
\end{table}

We examine the finite-sample performance of $\hat\psi$ and
$\tilde\psi$ for $p = 2$, $q \in \{10, 20, 30\}$, $\lambda = 8$ and
$n \in \{256, 512, 1024\}$. Treatment is assigned by complete
randomization, with $n_1 \in \{n / 16, 3 n /32, n / 8 \}$ treated,
independently of the covariates and errors. The two components of
$\bx_i= (x_{i1}, x_{i2})^\top$ are generated independently, with
$x_{i1}$ taking values zero and one with equal probability and
$x_{i2}$ following an exponential distribution with mean one.  We set
$\bmu = \b0_q$, $\alpha = 0$, $\gamma = 0.5$,
$\bdelta = (0.5, -0.5)^\top$, $\sigma^2 = 0.25$, $\Xi_{j1} = 0.4$,
$\Xi_{j2}=0.3(-1)^{j+1}$ and $\eta_j = 0$ $(j = 1, \ldots, q)$.  The
matrix $\bL$ is the lower Cholesky factor of the matrix with $(j,k)$
entry $2^{-|j-k|}$. The parameter has dimension $(p + 4)q + p +3$,
giving $65$, $125$, and $185$ estimated parameters, for $q = 10$,
$q = 20$, and $q = 30$, respectively.  We let $r = q / 2$ and
$k =q / 10$. For $\tau \in\{1,4\}$, we set
$(\nu_j, \beta_j) = (0.15, 0.125 \tau / r)$ for $j = 1, \ldots, r$,
$(\nu_j, \beta_j) = (0, 0)$ for $j = r + 1,\ldots,r + k$,
$(\nu_j, \beta_j) = (0, 0.125\tau / r)$ for
$j = r + k + 1, \ldots, r + 3k$, and $(\nu_j, \beta_j) = (0.15, 0)$
for $j = r + 3k + 1, \ldots, q$. In this way, half the mediators form
active pathways, and among the remainder, either treatment does not
affect the mediator, the mediator does not affect the outcome, or
both; Figure~\ref{fig:multiple-mediator-graph-q10} in the
Supplementary Material document shows a representation of the
corresponding causal diagram when $q = 10$.

We simulate $10^5$ independent datasets for each combination of $n$,
$q$, $\tau$, and $n_1$, keeping $\psi = 0.01875\tau$ fixed across $n$
and $q$. Table~\ref{tab:multiple-mediator-c1} compares $\hat\psi$ and
$\tilde\psi$ in terms of underestimation probability, mean bias, root
mean squared error, and coverage of nominally $95\%$ Wald and HulC
intervals for $\tau = 1$ and $n_1 = n / 16$. Median bias correction
brings the underestimation probability substantially closer to $0.5$
and reduces mean bias and root mean squared error.  Among the
intervals considered, only HulC intervals based on $\tilde\psi$ have
coverage consistently close to the nominal level.  The HulC intervals
based on $\hat\psi$ generally undercover, sometimes substantially, and
Wald intervals tend to
overcover. Section~\ref{sec:supp-multiple-mediator-results} of the
Supplementary Material provides results for the other combinations of
$\tau$ and $n_1$.

\end{example}

Section~\ref{sec:supp:marginal-effects-cont} of the Supplementary
Material document revisits Example~\ref{ex:marginal-effects} on
marginal effects, and illustrates how the MBC estimator enables HulC
to deliver intervals with near-nominal finite-sample coverage; in that
setting Wald-type inference, despite being the default in popular
computational frameworks, like the \texttt{marginaleffects} R package,
is typically unreliable even in situations with seemingly abundant
information about the parameters of generalized linear models (2500
observations for 6 parameters in Example~\ref{ex:marginal-effects}).

\subsection{Modified signed likelihood root}
\label{sec:rstar}

An approximately median-unbiased estimator $\psi^{*}$ can also be
obtained by solving $r^*(\psi) = 0$, where $r^*(\psi)$ is the modified
signed likelihood root \citep{barndorff-nielsen:1986}. The associated
confidence interval is obtained by equating $r^*(\psi)$ to standard
normal quantiles. Computing $r^*(\psi)$ requires sample-space
derivatives or approximations to them \citep{pierce+bellio:2017}, and
numerical inversion requires care near $\hat\psi$. For general models,
\citet{skovgaard:1996} proposed an approximation of the required
sample-space derivatives based on covariances of likelihood
quantities, which does not require conditioning on an ancillary
statistic. A practical implementation of that approach is provided by
the \texttt{likelihoodAsy} R package
\citep{likelihoodAsy}. Nevertheless, computation relies on repeated
simulation for estimating the required expectations, and the
evaluation of the constrained ML estimator $\hat\btheta_\psi$,
obtained under the restriction $h(\btheta)=\psi$, for a grid of values
of $\psi$. Apart from costly, constrained optimization can be unstable
or error-prone with small sample sizes, large $p$, or focus parameters
that are highly nonlinear functions of the parameters. In addition,
such methods cannot handle situations where $\hat\btheta$ or
$\hat\btheta_\psi$ lies on the boundary of the parameter space
\citep[see, for example,][for separation in multinomial logistic
regression]{albert+anderson:1984}.

Section~\ref{sec:supp:rstar} of the Supplementary Material document
provides a more detailed description of the modified signed likelihood
root and its approximations.  Section~\ref{sec:supp:weibull-example}
compares the MBC estimator with $\psi^{*}$ for a Weibull quantile, and
HulC intervals computed using MBC estimators with confidence intervals
obtained by inverting $r^*(\psi)$. In that example, the two estimators
achieve comparable control of median bias and both interval procedures
attain near-nominal coverage, while HulC intervals using MBC
estimators are substantially less costly to compute.


\section{Simulation-based focused median bias reduction}
\label{sec:simulation-based-focused}

As detailed in Section~\ref{sec:focused}, the
estimator~(\ref{eq:focused}) is readily available for a range of
widely-used models where the quantities $\iinfo(\btheta)$,
$\bP_t(\btheta)$ and $\bQ_t(\btheta)$ have been derived in closed
form. For general models, however, deriving those quantities may be
algebraically cumbersome and their implementation can be demanding,
even when likelihood evaluation and simulation from the fitted model
are straightforward; see, for example
\citet[Section~2.3]{grun+etal:2012} for an example of how involved the
expressions required for mean bias reduction in beta regression models
can be.
In such cases, focused median bias reduction can still be carried out
by estimating the required quantities by Monte Carlo simulation under
the model. This is one of the approaches that the accompanying
\texttt{focuson} \citep{focuson} R package implements.

Let $\btheta^\circ$ denote the value of the parameter at which the
correction is to be computed; for example,
$\btheta^\circ = \hat\btheta$ when the initial estimator is ML, or
$\btheta^\circ = \btheta^\dagger$ when a reduced mean-bias estimator
is used. Suppose that we can evaluate the log-likelihood
$l(\btheta; \bb y)$ at arbitrary $\btheta$ and data $\bb y$, and that
we can simulate independent samples $\by^{(1)}, \ldots, \by^{(R)}$
from $\mathcal{M}_{\btheta^\circ}$. For the $r$th simulated dataset
define $\bb{s}^{(r)} = \nabla l(\btheta^\circ; \bY^{(r)})$ and
$\bb{j}^{(r)} = -\nabla\nabla^\top l(\btheta^\circ; \bY^{(r)})$.
If analytic expressions for the score vector or the observed
information matrix are available, they can be used directly; otherwise
they can be computed numerically from the log-likelihood, or the
observed information can be computed as minus the numerical Jacobian of
the score. The Monte Carlo estimators
\begin{equation}
  \label{eq:simu-based}
  \widehat{\iinfo} = \left\{ \frac{1}{R}\sum_{r=1}^R \bb{s}^{(r)}\bb{s}^{(r)\top} \right\}^{-1} \,, \quad
  \widehat{\bP}_t = \frac{1}{R}\sum_{r=1}^R \bb{s}^{(r)}\bb{s}^{(r)\top}s_t^{(r)} \,,  \quad
  \widehat{\bQ}_t = - \frac{1}{R}\sum_{r=1}^R \bb{j}^{(r)}s_t^{(r)} \quad (t = 1, \ldots, p)
\end{equation}
are then substituted for $\iinfo$, $\bP_t$ and $\bQ_t$ in the
expressions in Section~\ref{sec:focused}. Using the second Bartlett
identity, an equivalent alternative for the estimator of $\iinfo$
under the regularity conditions of Section~\ref{sec:focused} is the
inverse of $R^{-1}\sum_{r=1}^R \bb{j}^{(r)}$. The numerical error due
to simulation is distinct from the statistical error of the estimator,
and can be made negligible by increasing $R$. Under suitable moment
conditions, the Monte Carlo error in the simulation-based median
bias-corrected estimator is $O_p(n^{-1/2}R^{-1/2})$.  Taking $R$ of
order $n^2$ therefore makes this error $O_p(n^{-3/2})$.
The same approach can be used to compute mean bias-corrected
estimators of focus parameters, starting from either the ML estimator
or a mean bias-reduced estimator of $\btheta$
(see, for example, Example~\ref{ex:marginal-effects}),
since~\eqref{eq:simu-based}
provides simulation-based estimates for all the required
quantities. 

For a full exponential family in canonical parameterization, or when
the data consist of $n$ independent and identically distributed
observations, the estimation of $\einfo$, $\bP_1,\ldots,\bP_p$, and
$\bQ_1,\ldots,\bQ_p$ simplifies considerably. Those simplifications
are described and illustrated, along with the benefits in estimation
and inference they deliver, for ordinal superiority measures in
adjacent-category models \citep{agresti+kateri:2017} and for the
circular variance of the first angular component in a bivariate von
Mises sine model \citep{singh+hnizdo+demchuk:2002,
  mardia+frellsen:2012}, in Section~\ref{sec:supp:ordinal-superiority}
and Section~\ref{sec:supp:bvmsin-circular-variance} of the
Supplementary Material document, respectively. In the latter example,
the required likelihood quantities involve derivatives of a
normalising constant that has only an infinite-series
representation. Hence, closed-form expressions are cumbersome. In
contrast, the sole inputs for focused median bias reduction and
HulC-based confidence intervals are the log-likelihood function, the
focus parameter as a function of $\btheta$, and a function to simulate
from the model at given $\btheta$ values.


\section{FIC model selection and inference about focus parameters}
\label{sec:fic}

\subsection{Focused information criterion}

The Focused Information Criterion (FIC) \citep{claeskens+hjort:2003}
selects the candidate model that minimizes an estimate of the
asymptotic mean squared error of the ML estimator of a focus parameter
under local misspecification. The wide model leaves all parameters
free, whereas the narrow model fixes some parameters at prespecified
values. Each candidate model is obtained by freeing a subset of the
parameters fixed in the narrow model, possibly in groups. Let $\mA$
denote the wide model and $\hat\mS$ the model selected by FIC. The
example below compares focus estimators from the wide and selected
models. For selection-agnostic inference, we construct HulC intervals
using the MBC estimator from the wide model.
Section~\ref{sec:supp:fic} of the Supplementary Material document
gives the details of the candidate models and the criterion.

\begin{example}[FIC selection and risk differences in logistic regression]
  \label{ex:fic}
  We consider the data set in
  \citet[Section~1.6.2]{hosmer+lemeshow:1989} about factors associated
  with low birth weights in infants, which has also been considered as
  an example of FIC model selection in
  \citet[Example~6.1]{claeskens+hjort:2008}. The data set consists of
  189 births at a US hospital, and records whether the birth weight is
  below 2.5~kg, the mother's race (white, black, other); smoking
  status during pregnancy; age; weight at last menstrual period;
  history of premature labours; history of hypertension; uterine
  irritability, and number of physician visits (zero, one, or more
  than one) during the first trimester.  We focus on the three
  race-specific risk differences for low birth weight due to smoking
  during pregnancy, for mothers with race-average age and weight, no
  history of premature labours or hypertension, no uterine
  irritability, and one physician visit.

  As a wide model, we consider the logistic regression model for low
  birth weight in terms of all available covariates, and, as the
  narrow model, the logistic regression model with only race and
  smoking status as covariates. Taking all subsets of the six
  remaining terms, with the indicators for physician visits included
  or excluded together, gives $2^6=64$ candidate models. We simulate
  $10^4$ samples from the ML fit of the wide model, and for each
  sample we estimate the focus parameters using ML and median bias
  correction from the wide model, and ML on the FIC-selected model,
  and compute the corresponding $95\%$ Wald-type confidence intervals
  and the $95\%$ HulC interval based on the MBC estimator. The MBC
  estimator uses the reduced mean-bias estimator of \citet{firth:1993}
  as its initial estimator, which guarantees finiteness of the
  estimates. We also consider a version of the ML estimator of the
  FIC-selected model corrected for its local-misspecification bias
  under the selected model.  Following the proposal of
  \citet[Section~7.5.2]{claeskens+hjort:2008}, the corresponding Wald
  interval is computed using the estimated standard error from the
  wide model. Table~\ref{tab:fic} shows the results. Median bias
  reduction results in almost median unbiased estimates, and in a
  substantial improvement of Wald-type inference compared to the other
  Wald-type procedures. The HulC-type intervals with $\Delta = 0$
  based on the MBC estimator achieve almost nominal coverage, despite
  there being fewer than 3 observations per parameter in each batch.
  
  \begin{table}[t!]
\centering
\begin{talltblr}[         
caption={Comparison of the ML focus estimator from the wide model ($\hat\psi_{(\mA)}$), the ML focus estimator and its corrected version from the FIC-selected model ($\hat\psi_{(\hat{\mS})}$ and cor.~$\hat\psi_{(\hat{\mS})}$, respectively), and the MBC estimator from the wide model ($\tilde\psi_{(\mA)}$) in terms of simulation-based estimates (see Example~\ref{ex:fic}) of probability of underestimation (PU), mean bias (BIAS), and root mean squared error (RMSE). The focus parameters are the three race-specific risk differences for low birth weight due to smoking during pregnancy. The column ``Wald'' reports the coverage of nominally $95\%$ Wald-type intervals. Those corresponding to the MBC estimator are based on supplied estimates of $\btheta$, and those based on cor.~$\hat\psi_{(\hat{\mS})}$ use the estimated standard error from the wide model.  The column ``HulC'' reports the coverage of nominally $95\%$ HulC-type intervals with $\Delta = 0$. All summaries are $\times 100$.},
label={tab:fic},
]                     
{                     
colspec={Q[]Q[]Q[]Q[]Q[]Q[]Q[]},
hline{2}={1-7}{solid, black, 0.05em},
hline{6}={1-7}{solid, black, 0.05em},
hline{10}={1-7}{solid, black, 0.05em},
hline{1}={1-7}{solid, black, 0.08em},
hline{14}={1-7}{solid, black, 0.08em},
row{1}={}{halign=c},
cell{2-13}{2}={}{halign=l},
cell{2-13}{3}={}{halign=r},
cell{2-13}{4}={}{halign=r},
cell{2-13}{5}={}{halign=r},
cell{2-13}{6}={}{halign=r},
cell{2-13}{7}={}{halign=r},
cell{2,6,10}{1}={r=4}{halign=l, valign=h},
cell{3-5,7-9,11-13}{1}={}{halign=l},
}                     
Race & Estimator & PU & BIAS & RMSE & Wald & HulC \\
white & $\hat\psi_{(\mA)}$ & 57.2 & -0.1 & 4.9 & 88.8 &   \\
white & $\hat\psi_{(\hat{\mS})}$ & 43.7 & 2.1 & 6.6 & 84.3 &   \\
white & cor.~$\hat\psi_{(\hat{\mS})}$ & 55.4 & 0.1 & 4.9 & 89.7 &   \\
white & $\tilde\psi_{(\mA)}$ & 50.5 & 0.6 & 5.0 & 92.8 & 94.8 \\
black & $\hat\psi_{(\mA)}$ & 54.2 & 0.1 & 9.5 & 88.5 &   \\
black & $\hat\psi_{(\hat{\mS})}$ & 42.0 & 3.4 & 11.1 & 84.2 &   \\
black & cor.~$\hat\psi_{(\hat{\mS})}$ & 52.4 & 0.6 & 9.7 & 88.7 &   \\
black & $\tilde\psi_{(\mA)}$ & 50.4 & 0.8 & 9.5 & 90.3 & 95.2 \\
other & $\hat\psi_{(\mA)}$ & 51.9 & 0.8 & 9.8 & 88.3 &   \\
other & $\hat\psi_{(\hat{\mS})}$ & 39.0 & 4.4 & 11.8 & 82.0 &   \\
other & cor.~$\hat\psi_{(\hat{\mS})}$ & 50.1 & 1.4 & 10.1 & 88.1 &   \\
other & $\tilde\psi_{(\mA)}$ & 49.8 & 1.1 & 9.6 & 90.0 & 94.8 \\
\end{talltblr}
\end{table}

\end{example}


\section{Concluding remarks}

A further practical advantage of median bias correction concerns the
range of the focus parameter. Because the correction is explicit, the
resulting estimate can occasionally fall outside that range. Provided
that it is plausible to assume that the focus parameter is an interior
point of its range, such an estimate can be truncated at the relevant
boundary without changing the events of under- and
overestimation. Truncation therefore preserves the median bias
properties of the estimator exactly. This contrasts with mean bias
correction: truncation changes expectations and hence reduction in
mean bias is not necessarily preserved.

Unlike ML and implicit median bias reduction, the explicit estimator
is not exactly equivariant: its numerical value depends on the chosen
parameterization. Nevertheless, its third-order median unbiasedness is
preserved under smooth monotone transformations of the focus
parameter. This distinction also creates flexibility in practice,
because both the parameterization and the initial estimator can be
selected to ensure stability or simplify computation.


\section{Supplementary material}

The Supplementary Material document provides the proof of
Theorem~\ref{thm:oracle} and all supplementary derivations and results
cited above. Scripts to reproduce all numerical results and outputs in
that document and the main text are at
\url{https://github.com/ikosmidis/fmedbr-supplementary}. The
accompanying \texttt{focuson} R package \citep{focuson} provides
general implementations of median bias correction and associated
methods.

\section{Declaration}

For the purpose of open access, the authors have applied a Creative
Commons Attribution (CC BY) licence to any Author Accepted Manuscript
version arising from this submission.


\appendix

\bibliographystyle{jss2}
\bibliography{arxiv-v2-references.bib}

\includepdf[page=-]{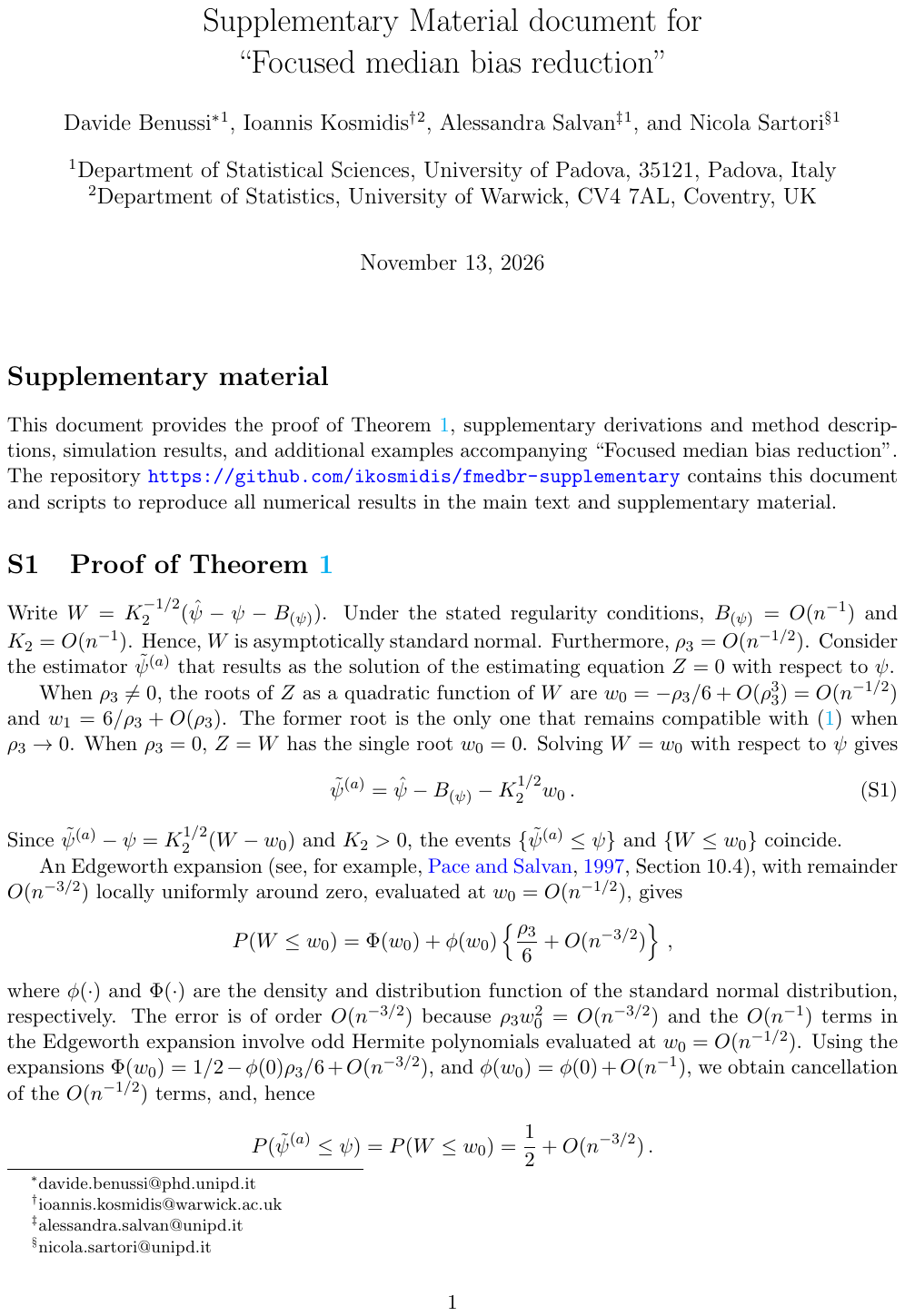}

\end{document}